\documentclass[aip, jmp, amsmath,amssymb, reprint,floatfix]{revtex4-1}

\usepackage{import}
\usepackage{graphicx}
\graphicspath{ {./Figures/} }
\usepackage{mathtools}
\usepackage{dcolumn}
\usepackage{bm}
\usepackage{braket}
\usepackage{enumitem}
\usepackage{float}
\usepackage[section]{placeins} 
\usepackage{booktabs}
\usepackage{array}
\usepackage{hyperref} 
\usepackage{verbatim}

\begin{document}

\title{Machine learning predictions of the Hessian matrix for peptides chains and small proteins}
\author{Giorgio Domenichini}

\begin{abstract} 
    Molecular Hessians have a key role in describing molecular vibrations, trajectories and optimization paths.  An explicit calculation of them through standard quantum mechanical methods can be computationally expensive for medium-large systems, and in many applications not even needed. Machine learning methods can be a shortcut to tackle efficiently the computational difficulties. This paper will present a ML model able to predict the Hessian matrix of biological system made of thousands of atoms. The method, based on learning the Hessian in internal coordinates is intrinsically invariant to molecular rotations and translations, and has a very good scaling with the systems' size. The training was performed on a dataset of simple aminoacids, as they constitute the building blocks of larger proteins. From the predicted Hessian matrix it is possible to calculate thermochemical properties within the harmonic approximations, among them enthalpies, entropies, Gibbs' free energies, and zero point vibrational energies.
    
\end{abstract}
\maketitle
\section{Introduction}

In the last two decades, machine learning (ML) has deeply impacted many aspects of computational chemistry.\cite{qml_ccs_anatole2018, qml_nutshell_rupp2015,von2020exploring, qml_rupp_atomization, faber_qml_lower_DFT, christensen2019operators, weinreich2021machine,  qml_properties, qml_optimization_hammer, qml_lemm2021energy, Keith2021QML} Many researches have been done in the topic of machine learning force fields, namely study the first derivative of the energy w.r.t. atoms' positions. \cite{mancini2020unsupervised,falbo2022integration,MLpotential_bartok2015} Other studies investigated the derivative of the energy w.r.t. the atomic charges. \cite{Levy1978, shiraogawa2023optimization,Keith2019,Keith2020, Griego2020MLcAP_Keith,Griego_Keith2021_comp_guidelines,von2007alchemical, Chang_bonds, Chang2018, Lilienfeld2009,anm, Domenichini2020, domenichini2022alchemical, domenichini2024acbs}

The molecular Hessian matrix contains the second derivatives of the molecular energy w.r.t atomic positions. The Hessian is an important molecular quantity because describes the curvature of the potential energy surface (PES), the knowledge of which is needed for geometrical optimization\cite{geomopt}, transition path sampling\cite{pathsampling}, calculation of molecular vibrations\cite{wilson1941} and vibrational thermochemical properties.\cite{grimme_free_energy} 

If in principle the Hessian matrix can be computed differentiating the machine learning forces with respect to the atomic positions.\cite{lam2020combining,review_ML_vib}
This process however requires the machine leaning force field to be very accurate and smoothly differentiable in the direction of all possible distortions. This can be only achieved by training the model on a large training set which includes many conformations of the same system.\cite{rupp_2022,MLpotential_Chmiela2017} So far only few papers propose new methods to predict directly the Hessian matrix \cite{review_han, han2025realtimeinterpretationneutronvibrational, dataNet, hessian_ml} or the vibrational frequencies. \cite{infrared_predictions,Wu2010_CFD,ceotto2013_CFD,Zhuang_2013_CFD}

The algorithm proposed in this paper is derived from one previously published\cite{domenichini2023molecular}. It can be summarized in three main steps. 
At first a set of redundant ICs is built from the geometry ad the connectivity of the molecule, according to the criteria outlined by Schlegel and coworkers. \cite{Peng,schlegel1982_Optimization,schlegel1984_hess_est}
The Hessian is than predicted in the IC system element by element using a local representations for every coordinate. 
In this paper will be used two ML models, a Random Forest Regression (RFR) (as in the previous paper), and a Neural Network (NN) regression model which can produce a faster estimate of the diagonal elements of the Internal Coordinate (IC) Hessian.
Finally the internal coordinate Hessian ($H_{IC}$) is transformed in Cartesian coordinates (CC) ($H_{CC}$) using the formalism proposed by Wilson\cite{wilson1941, wilson1955molecular}, vibrational properties can be computed from $H_{CC}$.

The training was performed on a newly built dataset which includes the 20 proteinogenic aminoacids in their neutral form, calculation were performed using the GFN2-xTB tight binding method. In this paper will be shown that it is possible to use aminoacids as building blocks  to predict the Hessian matrix for a series of peptides, and that the method can be applied to larger and biologically relevant systems. 

\section{Methods}
\subsection{Redundant internal Coordinates} \label{sect:RICs}
\subsubsection{Internal coordinates' definition} \label{sect:RIC_defin}
Redundant internal coordinates (also called primitive internal coordinates) were introduced by Schlegel and coworkers\cite{schlegel1982_Optimization,schlegel1984_hess_est,Geom_opt_largemols_Schegel,schlegel2011geometry} as a way to describe molecular geometries in terms of bonds, bond angles (after will be addressed as "angles"), and dihedral angles (after will be addressed as "dihedrals").
The construction of the internal coordinates, as well as the transformation of the Hessian from RICs to CCs and vice versa will be performed through using functions of the program geomeTRIC\cite{Geometric_ICs}. The algorithm works as follows.
At first the connectivity of the molecule is determined; a bond is defined between two atoms if the distance between them is shorter than 1.2 times the sum of their covalent radii (values are reported in literature\cite{Cordero_radii}).
Consequently, are defined all angles among three consecutively connected atoms, and all dihedrals among four connected angles.
Planar systems, like an sp$^2$ hybridized carbon atom ($I$) bonded to three atoms ($J$,$K$,$L$), where all four atoms lie on a plane, are treated differently. 
In order to reduce redundancy, only two of the three angles with $I$ as vertex are included in the IC set; in this paper a distinction will be made between "planar angles" and other bond angles.
To better describe atoms' displacements outside the plane, geomeTRIC adds to the ICs the "out of plane" dihedral angle ($I-J-K-L$).

Another special case is made for systems where three atoms lie on a line, in systems such conjugated dienes ($I=j=K$). Even if it does not occur in this work it is worth to recall the procedure. The angle $I=j=K$ is replaced with a pair of "Linear Angles", which measures the displacement of the central atom $J$ from the $I-K$ axis in the two orthogonal directions.
Any dihedral which includes the linear angle ($I=J=K-L$) it is not included in the ICs, however, these dihedrals are replaced with dihedrals which stretch through 5 atoms, built as if $I$ and $K$ where bonded directly (e.g. $L-I==K-M$).
 
\subsubsection{Coordinate transformations}
The transformations of the Hessians from Cartesian coordinates to Internal coordinates and vice versa is performed by geomeTRIC following the formalism of Pulay, Fogarasi and Schlegel \cite{fogarasi1992calculation,pulay1992geometry,Peng,schlegel_1998_vibr_analisis,allen1993_coord_transformations} the key points of which will be here recalled. 

The Wilson $B$ matrix,\cite{wilson1941, wilson1955molecular} is the matrix that condenses all partial derivatives of the internal coordinates $q_i$, with respect to the Cartesian coordinates $x_j$. For $B$, we define a right pseudo-inverse $B_{\rm inv}$. 
\begin{equation}
    \begin{aligned}
    	B_{ij} :&=\frac{\partial q_i}{\partial x_j}. \\
    	B_{\rm inv} &= B^T (B B^T)^{-1}
    \end{aligned}  
\end{equation}
The gradient can be transformed from internal coordinates  to Cartesian coordinates and vice versa, by multiplying it by $B^T $ and $B_{\rm inv}^T$ respectively.
\begin{equation}
	\begin{aligned}
    \mathbf{g}_{CC}&=B^T \mathbf{g}_{IC}.\\
    \mathbf{g}_{IC}&=B_{\rm inv}^T \mathbf{g}_{CC}. 
    \end{aligned}
\end{equation}
$B'$ is the derivative of $B$ with respect to the Cartesian displacement $x_k$. 
\begin{equation} \label{eq:Bpdef}
  B'_{ijk}:=\partial^2 q_i /\partial x_j \partial x_k,
\end{equation}

The Hessian can be transformed from internal to Cartesian coordinates and vice versa using the following transformations.
\begin{equation}
    \label{eq:hess_transformations}
    \begin{aligned}
    H_{CC}=B^T H_{IC} B+(B')^T\mathbf{g}_{IC} \\
    H_{IC}=B_{inv}^T (H_{CC}-(B')^T\mathbf{g}_{IC})B_{\rm inv}.
	\end{aligned}
\end{equation}

\subsection{Representations}   
 \label{sec:Representations}
In this work the RIC Hessian\ref{eq:RIC_Hess_def} will be predicted element by element for every pair of internal coordinates $q_i$ and $q_j$.
\begin{equation}
	\label{eq:RIC_Hess_def}
	(H_{IC})_{ij}:=\frac{\partial E}{\partial q_i \partial q_j}
\end{equation} 

The representations are built depending on the type of coordinates of $q_i$ and $q_j$, they can be either bonds, angles, dihedrals, planar angles, or out of plane distortions. Furthermore the coordinates are also classified based on their atoms types, this means that different models are trained to predict (as an example) C-C bonds, or C-H bonds, or any other type of bonds present inside the training set.

In the representations are contained local descriptors: nuclear charges, bond lengths, angles' and dihedrals' widths. Differently to our earlier work\cite{domenichini2023molecular} atomic bond orders will not be included in the representations.

Inside the representations nuclear charges are reported in atomic units ($e$), bond lengths as the reciprocal of their values measured in Bohr radii ($1/a_0$), angles and dihedrals widths ($\alpha$) are included through using the continuous periodic expression $1+\text{cos}(\alpha)$.

\subsubsection{Diagonal elements}
The diagonal elements of the RIC Hessian\ref{eq:RIC_Hess_def}, are the second derivatives of the Energy with respect to an internal coordinate $q_i==q_j$. 
Representations are built for every internal coordinate $q_i$ following the concepts introduced in the previous paper \cite{domenichini2023molecular}.

At first we need to identify the type of coordinate of $q_i$, it can be either a bond, an angle, a planar angle, dihedral or an out of plane distortion; the functions used to build the representations depend on the type of coordinate.

Bonds, angles, and dihedrals are further classified based on the nuclear charges of the atoms that define them. To give a unique classification, independent of atom indexing, the nuclear charges are listed in the order which places at first the highest charge.

Planar angle and out of plane distortion are also classified based on the nuclear charge of the $sp^2$ atom at the center of the planar system.

The representation vector itself contains a list of nuclear charges, bond lengths, angles' and dihedrals' widths, defined by the coordinate's edges and their closest neighbors. 

Depending on the type of coordinate, the algorithm carefully places and sorts these values, to ensure the uniqueness, and the consistency of the representations, and their invariance w.r.t. atomic indexing. Additional details, and the full code used to generate the coordinate specific representations, can be found online at the GitHub repository.\cite{Git_HML}

\subsubsection{Non-diagonal elements} \label{sec:non_diag}
Non diagonal elements of the RIC Hessian\ref{eq:RIC_Hess_def}, are mixed second derivatives of the Energy with respect to a pair of internal coordinate $\frac{\partial^2 E}{\partial q_i \partial q_j }$, where $q_i\ne q_j $. 

In general non-diagonal terms are smaller than diagonal terms, and in the limit where the two coordinates  $q_i$ and $q_j$ are far apart, the mixed derivatives go to zero. We therefore limit the learning to only a subset of meaningful coordinate pairs that are geometrically close, and share one of more edges. 

In particular we train ML models to predict the mixed second derivatives of the energy with respect to :

\begin {itemize} 
\item Two bonds sharing one atom ($ij$ and $jk$)
\item An angle $ijk$ and the bonds on its sides ($ij$ and $jk$)
\item An angle $ijk$ and the bonds $jl$ which have one extreme on the angle's vertex 
\item An angle $ijk$ and the bonds ($il$ and $kl$) which share one atom with the angle 
\item Two angles that share one side ($ijk$ and $jkl$) 
\item Two angles that share the vertex ($ijk$ and $ljm$)
\item A dihedral $ijkl$ and the bonds between two external atoms ($ij$, $kl$)
\item A dihedral $ijkl$ and the bond between the inner atoms ($jk$)
\item A dihedral $ijkl$ and the angles $ijk$ and $jkl$ 
\item A dihedral $ijkl$ and the angles $mij$ (or $klm$), where $m$ is an atom bonded to $i$ (or $l$)
\item A dihedral $ijkl$ and the angles $ijm$ (or $mkl$), where $m$ is an atom bonded to $j$ (or $k$)
\item A dihedral $ijkl$ and the angles $mjk$ ($jkm$), where $m$ is an atom bonded to $j$ (or $k$)
\item An out of plane distortion $ijkl$ and the bonds connecting the central atom to the others $ij$,$ik$,$il$
\item An out of plane distortion $ijkl$ and the planar angles which have $i$ as vertex ($jik$ $jil$ $kil$) 
\end{itemize}
The coordinate selections' criteria valid for angles are also applied to planar angles, but different ML model are trained.

For each pair of coordinates mentioned above was built a representation which contains geometrical descriptors (charges, bonds lengths, angles' widths) inherent to both coordinates, several criteria are applied to the representations to ensure the uniqueness, consistency and invariance w.r.t. atomic indexing.

\subsection{Computational Details}
In order to test the method we implemented a Python\cite{van1995python,ipython} 3.13 code, which uses Numpy\cite {harris2020numpy}, Scipy\cite{2020SciPy-NMeth} and Matplotlib\cite{hunter2007matplotlib} libraries, where possible parallelism was introduced through multiprocessing library\cite{python_multiprocessing}. The code is published on GitHub \cite{Git_HML} as part of the 'Hessian Machine Learning' project.

To test the size-scalability of the method, we show that it is possible to predict the Hessian of small-medium size peptides from a new generated dataset of 20 proteinogenic amino acids, summarized in table \ref{AA_table} as isolated molecules in their charge neutral form with their aminoacids' terminals capped with ACE (acetyl), and NME (N-methyl) groups. 
Initial structures were built using the AmberTools 23\cite{ambertools2023,amber_overview,amber_25} suite, in particular using tleap and the ff19SB force field parameters.\cite{ff19sb_paper}

Hessians and vibrational properties were computed for the optimized structures at the GFN2-xTB semiempirical extended tight binding level of theory using the xTB program suit. \cite{xtb_program,xtb_gfn2_grimmme2017, xtb_gfn2_grimmme2019}

Random forest regression models \cite{RFR_Kang,RFR_Dutschmann,RFR_Breiman2001,RFR_MOLS_Faber,faber_qml_lower_DFT,RFR_SOLUB_Palmer,RFR_Ward_mols} were trained using SciKitLearn subroutines \cite{scikit-learn}.

The forests consist of 30 decision trees, each one pruned to a maximum depth of 30 splits, the trained was performed over all the monomers' dataset using bootstrap aggregating to randomize the tree's generations. 

Neural networks were built inside JAX machine learning framework\cite{jax2026github} using the Flax library \cite{flax2026github}, they comprehend the input layer, a 6 nodes dense hidden layer with ReLU activation, and the output layer.
 
NNs were optimized using the Adam optimizer implemented inside Optax\cite{optax2026github}, the learning rate parameter was set to 10$^{-3}$.

After the training, the NN parameters were saved in the hard drive, as Numpy arrays ({\rm  .npy}), which can be loaded faster than a standard Orbax\cite{orbax2026github} checkpoint.

\section{Numerical Results}

\subsection{Benchmarking RFR model on homopeptide chains}
\label{sec:homopeptides}
After training the RFR models, the method was tested on a series of homopeptide chains with lenghts varying from 2 to 8 repeating units of the same aminoacid. Some representative aminoacids were chosen: alanine (neutral), glutamic acid, lysine (basic), asparagine (ammidic), proline (cyclic), histidine (aromatic).

\begin{table}[ht]
	\begin{tabular}{l|c|c|c|c|c|c|c}
		AA & 2 & 3 & 4 & 5 & 6 & 7 & 8 \\ \hline
		ALA & 7.2\% & 7.5\% & 7.6\% & 7.7\% & 7.8\% & 7.8\% & 7.8\% \\
		HYS & 14.1\% & 14.4\% & 14.6\% & 14.6\% & 14.7\% & 14.9\% & 15.0\% \\
		LYN & 8.0\% & 8.5\% & 8.8\% & 8.7\% & 8.8\% & 8.8\% & 8.7\% \\
		ASN & 9.2\% & 8.6\% & 8.8\% & 8.9\% & 9.5\% & 8.8\% & 8.6\% \\
		PRO & 10.5\% & 10.6\% & 10.2\% & 11.0\% & 10.3\% & 10.4\% & 10.5\% \\
		GLH & 8.4\% & 8.4\% & 9.0\% & 8.3\% & 8.7\% & 9.2\% & 9.4\% \\
	\end{tabular}
	\caption{The RFR's prediction error for homopeptides with length of 2-8 aminoacids, for the Hessian matrix in Cartesian coordinates.}
	\label{tab:homopeptides_error}
\end{table}

Table \ref{tab:homopeptides_error} shows the percent errors made by the ML predictions, reported as the ratio between the norm of the error and the xTB computed Hessian matrices (  ${|H^{ML}-H^{xTB}|}/{|H^{xTB}|}*100$ ).
The errors made for the whole matrix are around 10\%, we can address the main cause for the first error to the omission of several non-diagonal terms in the IC Hessian, which are counted as zero. In fact the highest errors are made for the peptides that contain ring systems (poly-histidines and poly-prolines), a pair of coordinates inside the ring can be strongly connected even if they do not belong to the categories listed in section\ref{sec:non_diag}.

Another way to gauge the accuracy of an Hessian prediction is to look at the thermodynamical properties that can be computed from it; the  thermochemical analysis was performed for both the GFN2-xTB and the ML Hessians, using the "thermo" module of the xTB program. \cite{thermo_module,xtb_program,thermal_grimme}

\begin{figure}[H] 
	\includegraphics[width=\linewidth]{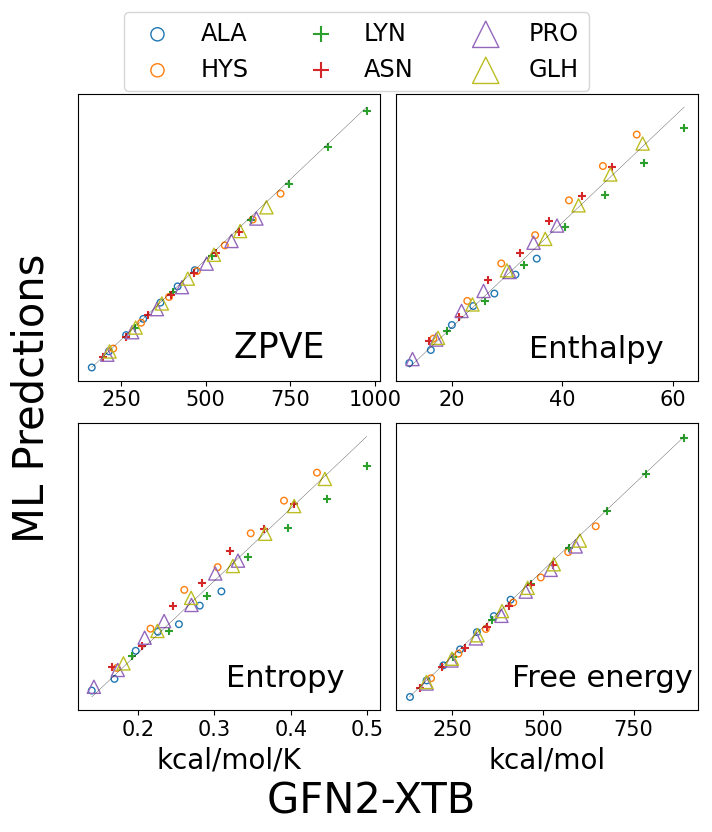}	\includegraphics[width=\linewidth]{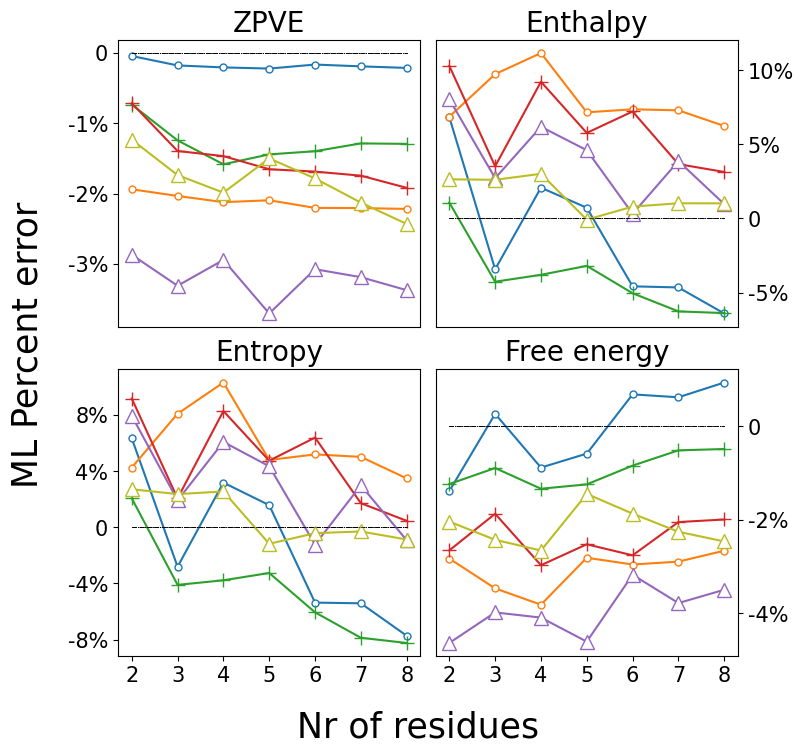}
	\caption{Thermochemical properties (for nuclear motions) of short peptide chains, calculated using the RRHO approach (at 298.15°K) for both the ML, and the GFN2-xTB Hessian. In the first panel are showed the values expressed in kcal/mol, or kcal/mol/K (for the entropy). In the second panel are showed the percent errors of the ML Hessian's results, compared to the GFN2-xTB values. }
	\label{fig:homopeptides}
\end{figure} 

In figure \ref{fig:homopeptides} are reported the zero point vibrational energies (ZPVE), and the total contributions of nuclear motions (translations+rotations+vibrations) to enthalpy ($H_{TOT}$), entropy ($S_{TOT}$), and Gibbs' free energy ($G_{TOT}$), calculated at the temperature $T=298.15°K$.

The ZPVE prediction error an underestimation of true value for all cases, this reflects the omission of some elements of the Hessian, thus it is possible to make similar comments to the ones made Table \ref{tab:homopeptides_error}.
The best results are made for the alanines, which are the most rigid systems, and the worst are made for the aromatic systems of the histidines and of the prolines, the error is in any case in the lower than 4\%, and in the most cases lower than 2\%.

For enthalpy and entropy the error is no more than 10\%, even if they are distinct quantities we can notice that the percent error is very similar for both quantities, this can be justified from the relations $H= -\partial ln(q)/\partial{\beta}$, and $S(T)=(U(T)-U(0))/T +R ln(q)$, and the calculation about how the error propagates from the Hessian, to the logarithm of the partition functions, and finally to enthalpy and entropy. 
We also can identify a downwards trend in the percent error, this can be justified by the fact that low frequency vibrations give the most important contributions to entropy and enthalpy. As the aminoacids chains get longer more low frequency modes arise, which can only partially described by the local ML model.

The Gibbs' free energy, which is the most important quantity directing chemical equilibrium, can be calculated from the previous three quantities from the relationship $G_{TOT}=ZPVE+H_{TOT}-TS_{TOT}$. 
The mean absolute percent error for the Gibbs' free energy is 2.2\%, with a maximum error of 4.6\%.
\subsection{Neural networks model}
\begin{figure}[H] 
	\centering
	\includegraphics[width=.4\linewidth]{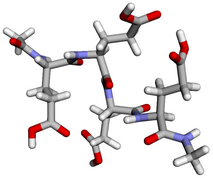}
	\hspace{.3cm}
	\includegraphics[width=.4\linewidth]{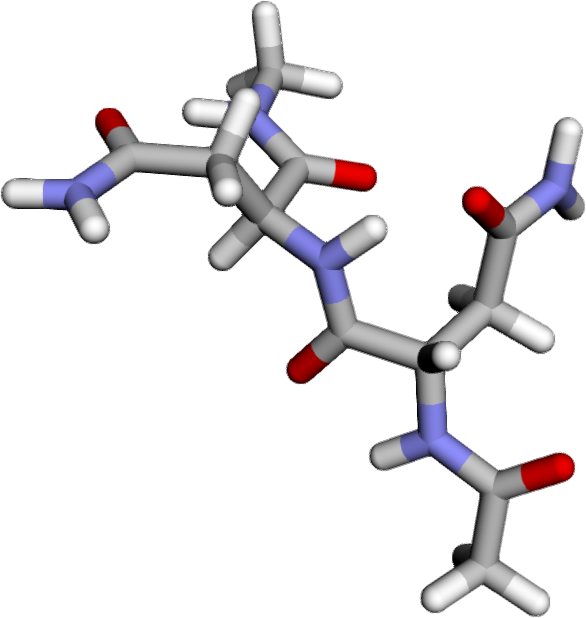}
	\caption{Peptide chains made of 4 glutamic acid, and 2 asparagine residues respectively.}
	\label{fig:GLH4_ASN2}
\end{figure}

As an alternative to the RFR method I also propose a lighter model to predict only the diagonal elements of the IC Hessian. Neural networks, implemented using the JAX library are in general faster to run and lighter to store than a RFR model, however the training becomes more long and difficult for non continuous representations such as the ones used for the non diagonal terms of the IC Hessian.
The model was trained to predict only the diagonal elements of the IC Hessian matrix, and was tested on the same homopeptides on which was tested the RFR model.
\begin{table}[ht]
	\begin{tabular}{l|c|c|c|c|c|c|c}
	& 2 & 3 & 4 & 5 & 6 & 7 & 8 \\ \hline
ALA & 7.8\% & 8.3\% & 8.6\% & 8.7\% & 8.9\% & 9.1\% & 9.1\% \\
HYS & 18.0\% & 18.5\% & 18.5\% & 18.5\% & 18.5\% & 18.5\% & 18.5\% \\
LYN & 7.6\% & 12.7\% & 12.7\% & 11.2\% & 11.0\% & 12.5\% & 11.0\% \\
ASN & 24.2\% & 11.9\% & 19.8\% & 12.7\% & 24.3\% & 12.9\% & 12.1\% \\
PRO & 17.0\% & 17.3\% & 16.0\% & 18.3\% & 15.7\% & 15.5\% & 15.1\% \\
GLH & 8.6\% & 14.2\% & 26.5\% & 9.2\% & 20.6\% & 38.1\% & 27.5\% \\
	\end{tabular}
	\caption{The NNs' prediction error for homopeptides with length of 2-8 aminoacids, for the diagonal elements of the IC Hessian matrix}
	\label{tab:homopeptides_errorNN}
\end{table}
Table \ref{tab:homopeptides_errorNN} shows the prediction error of the model, which is around 10\% for most systems.
A large error was found for the glutamic acid tetramer and eptamer, and the asparagine dimer, tetramer and hexamer. In these systems are present several intramolecular hydrogen bonds(as depicted, for example, in figure \ref{fig:GLH4_ASN2}).
The hydrogen bonds give more rigidity to the structure, thus many diagonal terms of the Hessian relative to angle and dihedral torsions are substantially bigger than their counterparts in the monomers' training set.

\subsection{Timing and scaling of the model}
\label{sec:timing}
In order to address the timing, and the scaling of both methods were chosen as a test systems a series of alanines peptide chains whose lengths vary from 10 up to 100 ALA residues (with an interval of 10); the number of atoms ranges from 112 to 1012, (every ALA residue has 10 atoms, while the ACE-NME cappings have 12 atoms).

The geometries were obtained using tleap and the force-field ff19SB , and not further optimized.  In this and in the following section the xTB gradient it is not computed, and assumed to be zero. 
This timings were obtained on a Linux system using a laptop with a Intel Core i7-4810MQ CPU and 31Gb of RAM, where possible many parts of the program where parallelized over 4 core. 

\begin{figure}[h]
	\includegraphics[width=.96\linewidth]{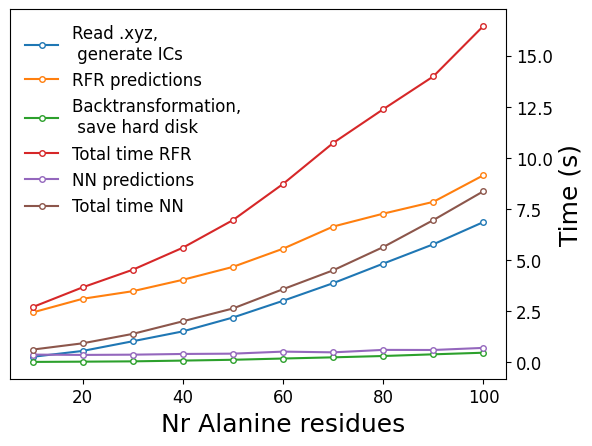}
	\caption{Timing the predictions of the NN, and the RFR models, the routine is divided in three parts (reading geometry input and generating the ICs, constructing the IC representation and predicting $H_{\rm IC}$, transforming $H_{\rm IC} \rightarrow H_{\rm CC}$ and saving it on the hard drive).}
	\label{fig:timing}
\end{figure}
Figure \ref{fig:timing} shows the timing of the method; to give a better insight, the routine is divided in three steps.
The first step consist of reading of the .xyz structures, and the construction of the internal coordinates. This step is performed using functions from the program "GeomeTRIC",  
we can see that it is very efficient for small and medium size systems, it takes less than 1 sec to build the ICs of peptides shorter than 30 alanine residues (312 atoms), however the computational burden increase approximately with $N^(3/2)$ with the system size, and the elapsed time becomes relevant for the predictions of very large systems.

The second step is the actual ML prediction of the IC Hessian, which consist in the construction of the representation, the loading of the saved models, the predictions of each individual Hessian matrix, and the reconstruction of the Hessian matrix itself. The most computationally expensive tasks are actually 
loading the models, and constructing of the representations.
The first task does not depend on the size of the system, rather on the number of models have to be loaded, namely the number different types of internal coordinates with different elements on the edges. Therefore this step is the slowest for small size systems, but does not affect the scalability of the method.
The RFR are stored as .joblib files, while the NN parameters are saved as numpy arrays, which are faster to load than Orbax checkpoints.
At the other end, building the representation is a task whose computational time grows linearly with the system size, and becomes the slowest routine for large systems. 
The transformation of the Hessian matrix from the internal to the Cartesian coordinate system follows Eq. \ref{eq:hess_transformations}. Inside the equation, the matrix product $B^T H_{IC} B$ can become very computationally expensive for systems with thousands of cartesian and internal coordinates. The most efficient way to compute the product for large systems, is by using the function \rm{scipy.sparse} to exploit the sparsity of the Wilson $B$ matrix.

\subsection{Biological examples} \label{sec:bio_examples}
In this section we would like to show how the method can scale up to some real system of biochemical interest. The structure of tetradotoxin was downloaded from PubChem\cite{pubchem_main, pubchem_TTX}, the structures of the other peptides were downloaded in .pdb format from the rcbs PDB data bank\cite{rcbs_paper,rcbs_url,wwpdb_paper,wwpdb_url} (identifiers: 1gCN\cite{1GCN_rcbs,1GCN_pdb,1GCN_ref} 3i40\cite{3i40_rcbs,3i40_pdb,3i40_ref}, 5o89\cite{5o89_rcbs,5o89_pdb,5o89_ref}, 5jQ3\cite{5jQ3_rcbs,5jQ3_pdb,5jQ3_ref}, 9dPF\cite{9dPF_rcbs,9dPF_pdb,9dPF_ref}).

\begin{figure}[H]
	\centering
	\includegraphics[width=.95\linewidth]{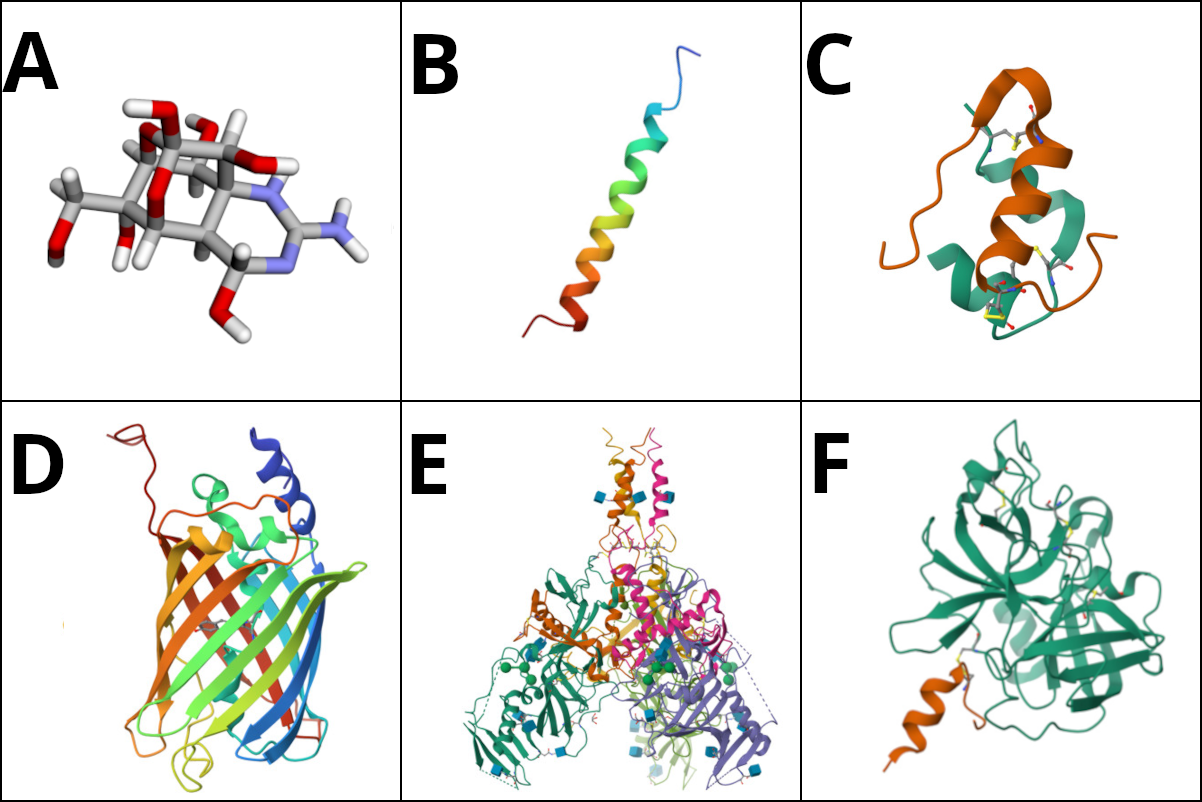}
	\caption{  (A) Tetradotoxin,(B) Glucagone, (C) Insulin, (D) rsEGFP2 ,  (E) Ebola virus glycoprotein  , (F) Transmembrane protease TMPRSS11D-S368A.}
	\label{fig:bio_examples}
\end{figure}

PDB files were manipulated using pdbfixer\cite{pdbfixer_github}(part of the OpenMM toolkit\cite{openmm7}) to add hydrogens and to remove crystallization water molecules, then converted to .xyz using openbabel\cite{openbabel_ref,openbabel_url}.   The molecular sketches and the ribbon plots in figure \ref{fig:bio_examples} were produced using py3DMol\cite{py3dmol} and mol*.\cite{rcbs_molstar}

The first molecule (A) is tetradotoxin, a deadly toxin which acts as an inhibitor of the sodium channels of the neural system.  
Tetradotoxin is produced by bacteria inside the body of several animal species, most notably inside the silver-cheeked puffer fish (\textit{Lagocephalus sceleratus}), an invasive alien species which recently caused an environmental emergency in the east Mediterranean sea.\cite{ttx1,ttx2,ttx3}

Glucagon (B) and Insulin (C) are well known peptide hormones which act as an antagonist pair while regulating the blood glucose levels. \cite{insuline_glucagone}

The reversibly photoswitchable enhanced green fluorescent protein (rsEGFP2) (D) is the most common mutant of the green fluorescent protein. It is widely used as a fluorescent biomarker due to his high fluorescence, small molecular mass, and its photoswitchable properties.\cite{gfp_storti,gfp2,rsegfp2_ref}

The Ebola virus glycoprotein (EBOV GP) (E), plays a crucial role in the attachment of the virus to the host cell, the endosomal entry and the membrane fusion.\cite{ebola1} Ebola virus is a deadly and infectious pathogen, which is responsible to many infections and fatalities occurred during several outbreaks.\cite{ebola2}

The human transmembrane protease (TMPRSS11D-S368A) (F),  together with other similar proteases, is involved in the mechanism of cell infection of several viruses, such as Rotaviruses\cite{tmprs1}, Influenza viruses\cite{tmprs2}, and SARS-CoV2.\cite{tmprs3}

\begin{table}[H]
	\begin{tabular}{l|c|c|c|c|c|c}
	& A& B & C &D&E&F \\ 
	Nr.residues & 1 & 29 & 51 & 243 & 384 & 498 \\
	Nr.atoms &39 & 471 & 782 & 3831 & 6025 & 7421 \\ \hline
	Generate ICs& 0.1" & 3" & 8" & 3'9" & 8'9" & 14'25" \\
	RFR model& 2" & 5" & 7" & 39" & 1'22" & 2'4" \\
	To H$_{\rm CC}$,save&0.0" & 0.1" & 0.3" & 10" & 40" & 1'0" \\
	Tot.time RFR& 2" & 9" & 16" & 3'59" & 10'11" & 17'30" \\  \hline
	Generate ICs & 0.1" & 3" & 8" & 3'11" & 8'44" & 15'19" \\
	NN model&0.0" & 0.3" & 0.2" & 3" & 8" & 12" \\
To H$_{\rm CC}$,save & 0.0" & 0.1" & 0.3" & 10" & 27" & 1'1" \\
	Tot.time NN&0.1" & 3" & 9" & 3'24" & 9'20" & 16'33" \\
	\end{tabular}	
	\label{tab:bio_examples}
\end{table}
In table\ref{tab:bio_examples} are summarized the computational times needed to compute the full Hessian matrix in Cartesian coordinates for the systems shown in Figure \ref{fig:bio_examples}. 
As in section \ref{sec:timing} the routine is divided in three main tasks: generating the internal coordinates from an .xyz file, use ML methods to give a prediction of the Hessian in internal coordinates, transform the Hessian from internal to Cartesian coordinates.

The full run of the program was performed twice, one time using RFR predictions, and one time using NN predictions. There is a small time difference in the first and third steps (ICs generation, and coordinate transformation)  of the two runs.

Many of the conclusions made in the previous section can also be confirmed from these timings. For large systems (D-E-F) the most computational demanding step is the generation of the ICs, and the transformation to Cartesian coordinates scales quadratically with the system size, however the machine learning models alone have a good scaling  with the system size. 

An explicit calculation of the Hessian even at xTB level can be prohibitive for systems of these size, while with the methods here proposed a ML approximation can be produced in a limited amount of time from any personal computer.

\section{Conclusions}

This paper presents a ML framework, based on a RFR model, to predict the Hessian matrix of a molecule, it derives from a previous work\cite{domenichini2023molecular} with some modifications to improve speed and versatility. 

It is also presented, as an alternative to RFR model, a NN model which can produce a fast prediction of the diagonal elements of the internal coordinate Hessian.

In Section\ref{sec:homopeptides} the models, trained on a set of aminoacids, are tested on a series of homopeptide chains. 
The RFR model tends to underestimate the Hessian matrix, with an error of approximately 10\%. However, this translates to a significantly lower error for the thermodynamical properties that can computed from the ML Hessian: the average error for the ZPVE the mean absolute percent error is 1.7\%.

The scaling of the model it is reported in section \ref{sec:timing}, the model can predict easily peptide chains with length up to 100 alanine residues, the prediction time for a system of 1012 atoms it is just 16.5 seconds.
It is also shown that the model can predict in a matter of minutes the Hessian matrix of some small proteins of up to 7400 atoms, which have an important role in the human metabolism, or in the transmission of pathogens.

Given the large dimension of this systems, the evaluation of the Hessian matrix by a quantum mechanical based method can be impossible.
From this work however the it is possible to obtain an accurate estimation of it, which can be sufficient in many application, e.g. as an initial Hessian guess in molecular dynamics simulations or in molecular geometry optimization.

The dataset on which the models were trained was based on GFN2-xTB calculation, which provided a good trade-off between accuracy and computing time, however a dataset larger, and built more carefully on DFT calculation can in principle produce more accurate prediction results at the same prediction times.

\section{Supplementary Material}
The supplementary materials of this paper contain two tables which report the full list of aminoacids used in the training, and the timings of the model on the alanine peptides. A theory section recalling the equations used inside the thermal analysis. 

\section{Data availability statement}

The data that support the findings of this study are available from the corresponding author upon reasonable request. The code developed to produce the results is available on GitHub \cite{Git_HML} as part of the 'Hessian Machine Learning' project.

\section{References}
\bibliography{Bib_Hessian}

@article{geomopt,
	author = {Schlegel, H. Bernhard},
	title = {Geometry optimization},
	journal = {WIREs Computational Molecular Science},
	volume = {1},
	number = {5},
	pages = {790-809},
	doi = {https://doi.org/10.1002/wcms.34},
	url = {https://wires.onlinelibrary.wiley.com/doi/abs/10.1002/wcms.34},
	year = {2011}
}

@article{pathsampling,
	author = {G. Bolhuis, Peter and Dellago, Christoph and Chandler, David},
	title = {Sampling ensembles of deterministic transition pathways },
	journal = {Faraday Discussions},
	volume = {110},
	pages = {421-436},
	year = {1998},
	month = {01},
	issn = {1359-6640},
	doi = {10.1039/a801266k},
	url = {https://doi.org/10.1039/a801266k}
}

@article{review_han,
	author = {Han, Bowen and Okabe, Ryotaro and Chotrattanapituk, Abhijatmedhi and Cheng, Mouyang and Li, Mingda and Cheng, Yongqiang},
	title = {AI-powered exploration of molecular vibrations, phonons, and spectroscopy},
	journal = {Digital Discovery},
	volume = {4},
	number = {3},
	pages = {584-624},
	year = {2025},
	month = {03},
    doi = {10.1039/d4dd00353e},
	url = {https://doi.org/10.1039/d4dd00353e},
}

@misc{han2025realtimeinterpretationneutronvibrational,
	title={Real-time interpretation of neutron vibrational spectra with symmetry-equivariant Hessian matrix prediction}, 
	author={Bowen Han and Pei Zhang and Kshitij Mehta and Massimiliano Lupo Pasini and Mingda Li and Yongqiang Cheng},
	year={2025},
	eprint={2502.13070},
	archivePrefix={arXiv},
	primaryClass={physics.chem-ph},
	url={https://arxiv.org/abs/2502.13070}, 
}

@article{review_ML_vib,
	author = {Han, Ruocheng and Ketkaew, Rangsiman and Luber, Sandra},
	title = {A Concise Review on Recent Developments of Machine
	Learning for the Prediction of Vibrational Spectra},
	journal = {The Journal of Physical Chemistry A},
	volume = {126},
	number = {6},
	pages = {801-812},
	year = {2022},
	month = {02},
	issn = {1089-5639},
	doi = {10.1021/acs.jpca.1c10417},
	url = {https://doi.org/10.1021/acs.jpca.1c10417}
}

@article{lam2020combining,
	title={Combining quantum mechanics and machine-learning calculations for anharmonic corrections to vibrational frequencies},
	author={Lam, Julien and Abdul-Al, Saleh and Allouche, Abdul-Rahman},
	journal={Journal of chemical theory and computation},
	volume={16},
	number={3},
	pages={1681--1689},
	year={2020},
	doi={https://doi.org/10.1021/acs.jctc.9b00964},
	publisher={ACS Publications}
}

@article{dataNet,
	title={A deep learning model for predicting selected organic molecular spectra},
	author={Zou, Zihan and Zhang, Yujin and Liang, Lijun and Wei, Mingzhi and Leng, Jiancai and Jiang, Jun and Luo, Yi and Hu, Wei},
	journal={Nature Computational Science},
	volume={3},
	number={11},
	pages={957--964},
	year={2023},
	publisher={Nature Publishing Group US New York},
	doi={https://doi.org/10.1038/s43588-023-00550-y}
}

@article{infrared_predictions,
	author = {Kartha, Adithya
	Ranjith and Ajayakumar, Dhanush P. and Idris, Muhammad and Ragupathy, Gopi},
	title = {Unlocking the Potential
	of Machine Learning in Enhancing
	Quantum Chemical Calculations for Infrared Spectral Prediction},
	journal = {ACS Omega},
	volume = {10},
	number = {18},
	pages = {19224-19234},
	year = {2025},
	month = {04},
	issn = {2470-1343},
	doi = {10.1021/acsomega.5c02405},
	url = {https://doi.org/10.1021/acsomega.5c02405},
}

@article{hessian_ml,
	author = {Gereon Feldmann  and Pit Steinbach  and Christoph Bannwarth },
	title = {A machine learning model for nuclear Hessians based on GFN2-xTB-derived atomistic features},
	journal = {ChemRxiv},
	volume = {2026},
	number = {0202},
	pages = {},
	year = {2026},
	doi = {10.26434/chemrxiv.10001864/v1},
	URL = {https://chemrxiv.org/doi/abs/10.26434/chemrxiv.10001864/v1}}

@article{Wu2010_CFD,
    author = {Wu, H. and Rahman, M. and Wang, J. and Louderaj, U. and Hase, W. L. and Zhuang, Y.},
    title = "{Higher-accuracy schemes for approximating the Hessian from electronic structure calculations in chemical dynamics simulations}",
    journal = {The Journal of Chemical Physics},
    volume = {133},
    number = {7},
    pages = {074101},
    year = {2010},
    month = {08},
    issn = {0021-9606},
    doi = {10.1063/1.3407922}
}

@article{ceotto2013_CFD,
    author = {Ceotto, Michele and Zhuang, Yu and Hase, William L.},
    title = {Accelerated direct semiclassical molecular dynamics using a compact finite difference Hessian scheme},
    journal = {The Journal of Chemical Physics},
    volume = {138},
    number = {5},
    pages = {054116},
    year = {2013},
    month = {02},
    issn = {0021-9606},
    doi = {10.1063/1.4789759}
}

@article{Zhuang_2013_CFD,
author = {Zhuang, Yu and Siebert, Matthew R. and Hase, William L. and Kay, Kenneth G. and Ceotto, Michele},
title = {Evaluating the Accuracy of Hessian Approximations for Direct Dynamics Simulations},
journal = {Journal of Chemical Theory and Computation},
volume = {9},
number = {1},
pages = {54-64},
year = {2013},
doi = {10.1021/ct300573h},
note ={PMID: 26589009},
URL = {https://doi.org/10.1021/ct300573h}
}

@article{MLpotential_bartok2015,
author = {Bartok, Albert P. and Csanyi, Gabor},
title = {Gaussian approximation potentials: A brief tutorial introduction},
journal = {International Journal of Quantum Chemistry},
volume = {115},
number = {16},
pages = {1051-1057},
doi = {https://doi.org/10.1002/qua.24927},
url = {https://onlinelibrary.wiley.com/doi/abs/10.1002/qua.24927},
eprint = {https://onlinelibrary.wiley.com/doi/pdf/10.1002/qua.24927},
year = {2015}
}

@article{MLpotential_Chmiela2017,
author = {Stefan Chmiela  and Alexandre Tkatchenko  and Huziel E. Sauceda  and Igor Poltavsky  and Kristof T. Schutt  and Klaus-Robert Muller },
title = {Machine learning of accurate energy-conserving molecular force fields},
journal = {Science Advances},
volume = {3},
number = {5},
pages = {e1603015},
year = {2017},
doi = {10.1126/sciadv.1603015}
}

@article{rupp_2022,
doi = {10.1088/2632-2153/aca005},
url = {https://dx.doi.org/10.1088/2632-2153/aca005},
year = {2022},
month = {nov},
publisher = {IOP Publishing},
volume = {3},
number = {4},
pages = {045017},
author = {Haoyan Huo and Matthias Rupp},
title = {Unified representation of molecules and crystals for machine learning},
journal = {Machine Learning: Science and Technology},
}

@misc{xtb_program,
  title={Semiempirical Extended Tight-Binding Program Package, xtb version 6.6.1},
  author={Ehlert, S and Bannwarth, C and Grimme, S},
  howpublished="\url{https://xtb-docs.readthedocs.io/en/latest/}"
}

@article{xtb_gfn2_grimmme2017,
author = {Grimme, Stefan and Bannwarth, Christoph and Shushkov, Philip},
title = {A Robust and Accurate Tight-Binding Quantum Chemical Method for Structures, Vibrational Frequencies, and Noncovalent Interactions of Large Molecular Systems Parametrized for All spd-Block Elements (Z=1-86)},
journal = {Journal of Chemical Theory and Computation},
volume = {13},
number = {5},
pages = {1989-2009},
year = {2017},
doi = {10.1021/acs.jctc.7b00118},
note ={PMID: 28418654},
URL = { https://doi.org/10.1021/acs.jctc.7b00118}
}

@article{xtb_gfn2_grimmme2019,
author = {Bannwarth, Christoph and Ehlert, Sebastian and Grimme, Stefan},
title = {GFN2-xTB—An Accurate and Broadly Parametrized Self-Consistent Tight-Binding Quantum Chemical Method with Multipole Electrostatics and Density-Dependent Dispersion Contributions},
journal = {Journal of Chemical Theory and Computation},
volume = {15},
number = {3},
pages = {1652-1671},
year = {2019},
doi = {10.1021/acs.jctc.8b01176},
note ={PMID: 30741547},
URL = {https://doi.org/10.1021/acs.jctc.8b01176}
}

@article{grimme_free_energy,
	author = {Spicher, Sebastian and Grimme, Stefan},
	title = {Efficient Computation of Free Energy Contributions
	for Association Reactions of Large Molecules},
	journal = {The Journal of Physical Chemistry Letters},
	volume = {11},
	number = {16},
	pages = {6606-6611},
	year = {2020},
	month = {07},
	issn = {1948-7185},
	doi = {10.1021/acs.jpclett.0c01930},
	url = {https://doi.org/10.1021/acs.jpclett.0c01930},
}

@article{entropy_grimme,
	author ={Pracht, Philipp and Grimme, Stefan},
	title  ={Calculation of absolute molecular entropies and heat capacities made simple},
	journal  ={Chem. Sci.},
	year  ={2021},
	volume  ={12},
	issue  ={19},
	pages  ={6551-6568},
	publisher  ={The Royal Society of Chemistry},
	doi  ={10.1039/D1SC00621E},
	url  ={http://dx.doi.org/10.1039/D1SC00621E}
	}

@article{thermal_grimme,
	author = {Grimme, Stefan},
	title = {Supramolecular Binding Thermodynamics by Dispersion-Corrected Density Functional Theory},
	journal = {Chemistry – A European Journal},
	volume = {18},
	number = {32},
	pages = {9955-9964},
	year={2012},
    doi = {https://doi.org/10.1002/chem.201200497},
	url = {https://chemistry-europe.onlinelibrary.wiley.com/doi/abs/10.1002/chem.201200497}
	}

@misc{thermo_module,
	howpublished="\url{https://xtb-docs.readthedocs.io/en/latest/xtb_thermo.html}"
}

@article{ff19sb_paper,
	author = {Tian, Chuan and Kasavajhala, Koushik and Belfon, Kellon A. A. and Raguette, Lauren and Huang, He and Migues, Angela N. and Bickel, John and Wang, Yuzhang and Pincay, Jorge and Wu, Qin and Simmerling, Carlos},
	title = {ff19SB: Amino-Acid-Specific Protein Backbone Parameters Trained against Quantum Mechanics Energy Surfaces in Solution},
	journal = {Journal of Chemical Theory and Computation},
	volume = {16},
	number = {1},
	pages = {528-552},
	year = {2020},
	doi = {10.1021/acs.jctc.9b00591},
	note ={PMID: 31714766},
	URL = {https://doi.org/10.1021/acs.jctc.9b00591}
	}

@article{ambertools2023,
	title={AmberTools},
	author={Case, David A and Aktulga, Hasan Metin and Belfon, Kellon and Cerutti, David S and Cisneros, G Andres and Cruzeiro, Vinicius Wilian D and Forouzesh, Negin and Giese, Timothy J and Gotz, Andreas W and Gohlke, Holger and others},
	journal={Journal of chemical information and modeling},
	volume={63},
	number={20},
	pages={6183},
	year={2023},
	doi={10.1021/acs.jcim.3c01153}
}

@article{amber_overview,
	author = {Salomon-Ferrer, Romelia and Case, David A. and Walker, Ross C.},
	title = {An overview of the Amber biomolecular simulation package},
	journal = {WIREs Computational Molecular Science},
	volume = {3},
	number = {2},
	pages = {198-210},
	doi = {https://doi.org/10.1002/wcms.1121},
	year = {2013}
}

@article{amber_25,
	author = {Case, David
	A. and Cerutti, David S. and Cruzeiro, Vinícius Wilian D. and Darden, Thomas A. and Duke, Robert E. and Ghazimirsaeed, Mahdieh and Giambasu, George M. and Giese, Timothy J. and Gotz, Andreas W. and Harris, Julie A. and Kasavajhala, Koushik and Lee, Tai-Sung and Li, Zhen and Lin, Charles and Liu, Jian and Miao, Yinglong and Salomon-Ferrrer, Romelia and Shen, Jana and Snyder, Ryan and Swails, Jason and Walker, Ross C. and Wang, Jinan and Wu, Xiongwu and Zeng, Jinzhe and Cheatham III, Thomas E. and Roe, Daniel R. and Roitberg, Adrian and Simmerling, Carlos and York, Darrin M. and Nagan, Maria C. and Merz, Kenneth M, Jr},
	title = {Recent Developments
	in Amber Biomolecular Simulations},
	journal = {Journal of Chemical Information and Modeling},
	volume = {65},
	number = {15},
	pages = {7835-7843},
	year = {2025},
	month = {07},
	issn = {1549-9596},
	doi = {10.1021/acs.jcim.5c01063},
}

@article{shiraogawa2023optimization,
    author = {Shiraogawa, Takafumi and Hasegawa, Jun-ya},
    title = {Optimization of
General Molecular Properties in the
Equilibrium Geometry Using Quantum Alchemy: An Inverse Molecular Design
Approach},
    journal = {The Journal of Physical Chemistry A},
    volume = {127},
    number = {19},
    pages = {4345-4353},
    year = {2023},
    month = {05},
    issn = {1089-5639},
    doi = {10.1021/acs.jpca.3c00205},
}

@article{Keith2019,
  title={Benchmarking computational alchemy for carbide, nitride, and oxide catalysts},
  author={Griego, Charles D and Saravanan, Karthikeyan and Keith, John A},
  journal={Advanced Theory and Simulations},
  volume={2},
  number={4},
  pages={1800142},
  year={2019},
  publisher={Wiley Online Library},
  doi={https://doi.org/10.1002/adts.201800142}
}

@article{Keith2020,
author = {Griego, Charles D. and Kitchin, John R. and Keith, John A.},
title = {Acceleration of catalyst discovery with easy, fast, and reproducible computational alchemy},
journal = {International Journal of Quantum Chemistry},
volume = {121},
number = {1},
pages = {e26380},
doi = {https://doi.org/10.1002/qua.26380},
url = {https://onlinelibrary.wiley.com/doi/abs/10.1002/qua.26380},
year = {2021}
}

@article{Griego2020MLcAP_Keith,
author = {Griego, Charles D. and Zhao, Lingyan and Saravanan, Karthikeyan and Keith, John A.},
title = {Machine learning corrected alchemical perturbation density functional theory for catalysis applications},
journal = {AIChE Journal},
volume = {66},
number = {12},
pages = {e17041},
doi = {https://doi.org/10.1002/aic.17041},
url = {https://aiche.onlinelibrary.wiley.com/doi/abs/10.1002/aic.17041},
year = {2020}
}

@article{Griego_Keith2021_comp_guidelines,
author = {Griego, Charles D. and Maldonado, Alex M. and Zhao, Lingyan and Zulueta, Barbaro and Gentry, Brian M. and Lipsman, Eli and Choi, Tae Hoon and Keith, John A.},
title = {Computationally Guided Searches for Efficient Catalysts through Chemical/Materials Space: Progress and Outlook},
journal = {The Journal of Physical Chemistry C},
volume = {125},
number = {12},
pages = {6495-6507},
year = {2021},
doi = {10.1021/acs.jpcc.0c11345},
URL = { https://doi.org/10.1021/acs.jpcc.0c11345}
}

@article{Keith2021QML,
author = {Keith, John A. and Vassilev-Galindo, Valentin and Cheng, Bingqing and Chmiela, Stefan and Gastegger, Michael and Muller, Klaus-Robert and Tkatchenko, Alexandre},
title = {Combining Machine Learning and Computational Chemistry for Predictive Insights Into Chemical Systems},
journal = {Chemical Reviews},
volume = {121},
number = {16},
pages = {9816-9872},
year = {2021},
doi = {10.1021/acs.chemrev.1c00107},
    note ={PMID: 34232033},
URL = { 
        https://doi.org/10.1021/acs.chemrev.1c00107
},
eprint = { https://doi.org/10.1021/acs.chemrev.1c00107}
}

@article{Levy1978,
author = {Levy,Mel },
title = {An energy‐density equation for isoelectronic changes in atoms},
journal = {J. Chem. Phys.},
volume = {68},
number = {11},
pages = {5298-5299},
year = {1978},
doi = {10.1063/1.435604},
URL = { https://doi.org/10.1063/1.435604},
eprint = {https://doi.org/10.1063/1.435604
}}

@article{von2007alchemical,
  title={Alchemical variations of intermolecular energies according to molecular grand-canonical ensemble density functional theory},
  author={Von Lilienfeld, O Anatole and Tuckerman, ME},
  journal={Journal of chemical theory and computation},
  volume={3},
  number={3},
  pages={1083--1090},
  year={2007},
  publisher={ACS Publications},
  doi={https://doi.org/10.1021/ct700002c}
}

@article{Chang_bonds,
author = {Chang,K. Y. Samuel  and Fias,Stijn  and Ramakrishnan,Raghunathan  and von Lilienfeld,O. Anatole },
title = {Fast and accurate predictions of covalent bonds in chemical space},
journal = {J. Chem. Phys.},
volume = {144},
number = {17},
pages = {174110},
year = {2016},
doi = {10.1063/1.4947217},
URL = { https://doi.org/10.1063/1.4947217  },
eprint = { https://doi.org/10.1063/1.4947217 }
}

@article{Chang2018,
  title = {$\text{Al}_x \text{Ga}_{1\ensuremath{-}x}\text{As}$ crystals with direct 2 eV band gaps from computational alchemy},
  author = {Chang, K. Y. Samuel and von Lilienfeld, O. Anatole},
  journal = {Phys. Rev. Materials},
  volume = {2},
  issue = {7},
  pages = {},
  numpages = {9},
  year = {2018},
  month = {Jul},
  publisher = {American Physical Society},
  doi = {10.1103/PhysRevMaterials.2.073802},
  url ={https://link.aps.org/doi/10.1103/PhysRevMaterials.2.073802}
}

@article{Lilienfeld2009,
    author = {O. Anatole von Lilienfeld},
    title = {Accurate ab initio energy gradients in chemical compound space},
    journal = {J. Chem. Phys.},
    volume = {131},
    number = {8},
    pages = {164102},
    year = {2009},
    doi = {https://doi.org/10.1063/1.3249969}
}

@article{anm,
author = {Fias, Stijn and Chang, K. Y. Samuel and von Lilienfeld, O. Anatole},
title = {Alchemical Normal Modes Unify Chemical Space},
journal = {J. Phys. Chem. Lett.},
volume = {10},
number = {1},
pages = {30-39},
year = {2019},
doi = {10.1021/acs.jpclett.8b02805},
URL = { https://doi.org/10.1021/acs.jpclett.8b02805}
}

@article{falbo2022integration,
  title={Integration of Quantum Chemistry, Statistical Mechanics, and Artificial Intelligence for Computational Spectroscopy: The UV--Vis Spectrum of TEMPO Radical in Different Solvents},
  author={Falbo, Emanuele and Fuse, Marco and Lazzari, Federico and Mancini, Giordano and Barone, Vincenzo},
  journal={Journal of Chemical Theory and Computation},
  volume={18},
  number={10},
  pages={6203--6216},
  year={2022},
  publisher={ACS Publications},
  doi={ https://doi.org/10.1021/acs.jctc.2c00654}
}

@article{mancini2020unsupervised,
  title={Unsupervised search of low-lying conformers with spectroscopic accuracy: A two-step algorithm rooted into the island model evolutionary algorithm},
  author={Mancini, Giordano and Fuse, Marco and Lazzari, Federico and Chandramouli, Balasubramanian and Barone, Vincenzo},
  journal={The Journal of Chemical Physics},
  volume={153},
  number={12},
  pages={124110},
  year={2020},
  publisher={AIP Publishing LLC},
  doi={ https://doi.org/10.1063/5.0018314}
}

@book{atkins2023atkins,
	title={Atkins' physical chemistry},
	author={Atkins, Peter William and De Paula, Julio and Keeler, James},
	year={2023},
	publisher={Oxford university press}
}

@article{Domenichini2020,
author = {Domenichini,Giorgio  and von Rudorff,Guido Falk  and von Lilienfeld,O. Anatole },
title = {Effects of perturbation order and basis set on alchemical predictions},
journal = {The Journal of Chemical Physics},
volume = {153},
number = {14},
pages = {144118},
year = {2020},
doi = {10.1063/5.0023590},
URL = {  https://doi.org/10.1063/5.0023590
},
eprint = { https://doi.org/10.1063/5.0023590}
}

@article{domenichini2022alchemical,
  title={Alchemical geometry relaxation},
  author={Domenichini, Giorgio and von Lilienfeld, O Anatole},
  journal={The Journal of Chemical Physics},
  volume={156},
  number={18},
  pages={184801},
  year={2022},
  publisher={AIP Publishing LLC},
  doi={ https://doi.org/10.1063/5.0085817}
}

@article{domenichini2024acbs,
	author = {Domenichini, Giorgio},
	title = {Extending the definition of atomic basis sets to atoms with fractional nuclear charge},
	journal = {The Journal of Chemical Physics},
	volume = {160},
	number = {12},
	pages = {124107},
	year = {2024},
	month = {03},
	issn = {0021-9606},
	doi = {10.1063/5.0196383},
	url = {https://doi.org/10.1063/5.0196383},
}

@article{domenichini2023molecular,
   author = {Domenichini, Giorgio and Dellago, Christoph},
title = {Molecular Hessian matrices from a machine learning random forest regression algorithm},
journal = {The Journal of Chemical Physics},
volume = {159},
number = {19},
pages = {194111},
year = {2023},
month = {11},
issn = {0021-9606},
doi =  {https://doi.org/10.1063/5.0169384}
}

@article{schlegel2011geometry,
  title={Geometry optimization},
  author={Schlegel, H Bernhard},
  journal={Wiley Interdisciplinary Reviews: Computational Molecular Science},
  volume={1},
  number={5},
  pages={790--809},
  year={2011},
  publisher={Wiley Online Library},
  doi={https://doi.org/10.1002/wcms.34}
}

@article{pulay1992geometry,
  title={Geometry optimization in redundant internal coordinates},
  author={Pulay, Peter and Fogarasi, Geza},
  journal={The Journal of chemical physics},
  volume={96},
  number={4},
  pages={2856--2860},
  year={1992},
  publisher={American Institute of Physics},
  doi={https://doi.org/10.1063/1.462844}
}

@article{fogarasi1992calculation,
  title={The calculation of ab initio molecular geometries: efficient optimization by natural internal coordinates and empirical correction by offset forces},
  author={Fogarasi, Geza and Zhou, Xuefeng and Taylor, Patterson W and Pulay, Peter},
  journal={Journal of the American Chemical Society},
  volume={114},
  number={21},
  pages={8191--8201},
  year={1992},
  publisher={ACS Publications},
  doi={ https://doi.org/10.1021/ja00047a032}
}

@article{wilson1941,
	author = {Wilson, E. Bright, Jr.},
	title = {Some Mathematical Methods for the Study of Molecular Vibrations},
	journal = {The Journal of Chemical Physics},
	volume = {9},
	number = {1},
	pages = {76-84},
	year = {1941},
	month = {01},
	issn = {0021-9606},
	doi = {10.1063/1.1750829}
}

@article{wilson1955molecular,
  title={Molecular vibrations},
  author={Wilson Jr, E Bright and Decius, John G and Cross, Paul G},
  journal={American Journal of Physics},
  volume={23},
  number={8},
  pages={550--550},
  year={1955},
  publisher={American Association of Physics Teachers},
  doi={10.1063/1.3061820}
}

@article{allen1993_coord_transformations,
  title={On the ab initio determination of higher-order force constants at nonstationary reference geometries},
  author={Allen, Wesley D and Csaszar, Attila G},
  journal={The Journal of chemical physics},
  volume={98},
  number={4},
  pages={2983--3015},
  year={1993},
  publisher={American Institute of Physics},
  doi={https://doi.org/10.1063/1.464127/},
}

@article{Born_Oppenheimer,
author = {Born, M. and Oppenheimer, R.},
title = {Zur Quantentheorie der Molekeln},
journal = {Annalen der Physik},
volume = {389},
number = {20},
pages = {457-484},
doi = {https://doi.org/10.1002/andp.19273892002},
url = {https://onlinelibrary.wiley.com/doi/abs/10.1002/andp.19273892002},
eprint = {https://onlinelibrary.wiley.com/doi/pdf/10.1002/andp.19273892002},
year = {1927}
}

@article{qml_ccs_anatole2018,
  title={Quantum machine learning in chemical compound space},
  author={Von Lilienfeld, O Anatole},
  journal={Angewandte Chemie International Edition},
  volume={57},
  number={16},
  pages={4164-4169},
  year={2018},
  publisher={Wiley Online Library},
  doi={https://doi.org/10.1002/anie.201709686}
}

@article{qml_nutshell_rupp2015,
  title={Machine learning for quantum mechanics in a nutshell},
  author={Rupp, Matthias},
  journal={International Journal of Quantum Chemistry},
  volume={115},
  number={16},
  pages={1058--1073},
  year={2015},
  publisher={Wiley Online Library},
  doi={https://doi.org/10.1002/qua.24954}
}

@article{qml_lemm2021energy,
  title={Machine learning based energy-free structure predictions of molecules, transition states, and solids},
  author={Lemm, Dominik and von Rudorff, Guido Falk and von Lilienfeld, O Anatole},
  journal={Nature Communications},
  volume={12},
  number={1},
  pages={1--10},
  year={2021},
  publisher={Nature Publishing Group}
}

@article{qml_optimization_hammer,
  title = {Global optimization of atomic structure enhanced by machine learning},
  author = {Bisbo, Malthe K. and Hammer, Bj\o{}rk},
  journal = {Phys. Rev. B},
  volume = {105},
  issue = {24},
  pages = {245404},
  numpages = {15},
  year = {2022},
  month = {Jun},
  publisher = {American Physical Society},
  doi = {10.1103/PhysRevB.105.245404},
  url = {https://link.aps.org/doi/10.1103/PhysRevB.105.245404}
}

@article{qml_rupp_atomization,
  title={Fast and accurate modeling of molecular atomization energies with machine learning},
  author={Rupp, Matthias and Tkatchenko, Alexandre and Muller, Klaus-Robert and Von Lilienfeld, O Anatole},
  journal={Phys. Rev. Lett.},
  volume={108},
  number={5},
  pages={058301},
  year={2012},
  publisher={APS},
doi={https://doi.org/10.1103/PhysRevLett.108.058301}
}

@article{faber_qml_lower_DFT,
  title={Prediction errors of molecular machine learning models lower than hybrid DFT error},
  author={Faber, Felix A and Hutchison, Luke and Huang, Bing and Gilmer, Justin and Schoenholz, Samuel S and Dahl, George E and Vinyals, Oriol and Kearnes, Steven and Riley, Patrick F and Von Lilienfeld, O Anatole},
  journal={J. Chem. Theory Comput.},
  doi={https://doi.org/10.1021/acs.jctc.7b00577},
  volume={13},
  number={11},
  pages={5255--5264},
  year={2017},
  publisher={ACS Publications},

}

@article{weinreich2021machine,
  title={Machine learning of free energies in chemical compound space using ensemble representations: Reaching experimental uncertainty for solvation},
  author={Weinreich, Jan and Browning, Nicholas J and von Lilienfeld, O Anatole},
  journal={The Journal of Chemical Physics},
  volume={154},
  number={13},
  pages={134113},
  year={2021},
  publisher={AIP Publishing LLC},
  doi={ https://doi.org/10.1063/5.0041548}
}

@article{christensen2019operators,
  title={Operators in quantum machine learning: Response properties in chemical space},
  author={Christensen, Anders S and Faber, Felix A and von Lilienfeld, O Anatole},
  journal={The Journal of chemical physics},
  volume={150},
  number={6},
  pages={064105},
  year={2019},
  publisher={AIP Publishing LLC},
  doi = {10.1063/1.5053562},

}

@article{qml_properties,
  title={Accelerating materials property predictions using machine learning},
  author={Pilania, Ghanshyam and Wang, Chenchen and Jiang, Xun and Rajasekaran, Sanguthevar and Ramprasad, Ramamurthy},
  journal={Scientific Reports},
  volume={3},
  number={1},
  pages={1--6},
  year={2013},
  publisher={Nature Publishing Group}
}

@article{von2020exploring,
  title={Exploring chemical compound space with quantum-based machine learning},
  author={von Lilienfeld, O Anatole and Muller, Klaus-Robert and Tkatchenko, Alexandre},
  journal={Nature Reviews Chemistry},
  volume={4},
  number={7},
  pages={347--358},
  year={2020},
  publisher={Nature Publishing Group},
  doi={https://doi.org/10.1038/s41570-020-0189-9}
}

@article{RFR_Breiman2001,
author={Breiman, Leo},
year= {2001},
title={Random Forests},
journal={Machine Learning},
pages={5-32},
volume={45},
issue={1},
doi={10.1023/A:1010933404324}
}

@article{RFR_Kang,
author = {Kang, Beomchang and Seok, Chaok and Lee, Juyong},
title = {Prediction of Molecular Electronic Transitions Using Random Forests},
journal = {Journal of Chemical Information and Modeling},
volume = {60},
number = {12},
pages = {5984-5994},
year = {2020},
doi = {10.1021/acs.jcim.0c00698},
note ={PMID: 33090804},
URL = { https://doi.org/10.1021/acs.jcim.0c00698},
eprint = { https://doi.org/10.1021/acs.jcim.0c00698}
}

@article{RFR_Dutschmann,
AUTHOR = {Dutschmann, Thomas-Martin and Baumann, Knut},
TITLE = {Evaluating High-Variance Leaves as Uncertainty Measure for Random Forest Regression},
JOURNAL = {Molecules},
VOLUME = {26},
YEAR = {2021},
NUMBER = {21},
ARTICLE-NUMBER = {6514},
URL = {https://www.mdpi.com/1420-3049/26/21/6514},
PubMedID = {34770921},
ISSN = {1420-3049},
DOI = {10.3390/molecules26216514}
}

@article{RFR_MOLS_Faber,
author = {Faber,Felix A.  and Christensen,Anders S.  and Huang,Bing  and von Lilienfeld,O. Anatole },
title = {Alchemical and structural distribution based representation for universal quantum machine learning},
journal = {The Journal of Chemical Physics},
volume = {148},
number = {24},
pages = {241717},
year = {2018},
doi = {10.1063/1.5020710},
URL = { https://doi.org/10.1063/1.5020710},
eprint = { https://doi.org/10.1063/1.5020710}
}

@article{RFR_Ward_mols,
  title = {Including crystal structure attributes in machine learning models of formation energies via Voronoi tessellations},
  author = {Ward, Logan and Liu, Ruoqian and Krishna, Amar and Hegde, Vinay I. and Agrawal, Ankit and Choudhary, Alok and Wolverton, Chris},
  journal = {Phys. Rev. B},
  volume = {96},
  issue = {2},
  pages = {024104},
  numpages = {12},
  year = {2017},
  month = {Jul},
  publisher = {American Physical Society},
  doi = {10.1103/PhysRevB.96.024104},
  url = {https://link.aps.org/doi/10.1103/PhysRevB.96.024104}
}

@article{RFR_SOLUB_Palmer,
author = {Palmer, David S. and O'Boyle, Noel M. and Glen, Robert C. and Mitchell, John B. O.},
title = {Random Forest Models To Predict Aqueous Solubility},
journal = {Journal of Chemical Information and Modeling},
volume = {47},
number = {1},
pages = {150-158},
year = {2007},
doi = {10.1021/ci060164k},
    note ={PMID: 17238260},
URL = { https://doi.org/10.1021/ci060164k},
eprint = { https://doi.org/10.1021/ci060164k}
}

@article{schlegel1982_Optimization,
author = {Schlegel, H. Bernhard},
title = {Optimization of equilibrium geometries and transition structures},
journal = {Journal of Computational Chemistry},
volume = {3},
number = {2},
pages = {214-218},
doi = {https://doi.org/10.1002/jcc.540030212},
url = {https://onlinelibrary.wiley.com/doi/abs/10.1002/jcc.540030212},
eprint = {https://onlinelibrary.wiley.com/doi/pdf/10.1002/jcc.540030212},
year = {1982}
}

@article{schlegel1984_hess_est,
author  = {Bernhard Schlegel, H.},
year=   {1984},
date= {1984/09/01},
title  = {Estimating the hessian for gradient-type geometry optimizations},
journal  = {Theoretica chimica acta},
pages  = {333-340},
volume  = {66},
issue  = {5},
SN  = {1432-2234},
url={ https://doi.org/10.1007/BF00554788},
doi={ 10.1007/BF00554788}
}

@article{Peng,
  title={Using redundant internal coordinates to optimize equilibrium geometries and transition states},
  author={Peng, Chunyang and Ayala, Philippe Y and Schlegel, H Bernhard and Frisch, Michael J},
  journal={Journal of Computational Chemistry},
  volume={17},
  number={1},
  pages={49--56},
  year={1996},
  publisher={Wiley Online Library},
  doi={https://doi.org/10.1002/(SICI)1096-987X(19960115)17:1<49::AID-JCC5>3.0.CO;2-0}
}

@article{Geom_opt_largemols_Schegel,
author = {Farkas,Odon  and Schlegel,H. Bernhard},
title = {Methods for geometry optimization of large molecules. I. An O(N2) algorithm for solving systems of linear equations for the transformation of coordinates and forces},
journal = {The Journal of Chemical Physics},
volume = {109},
number = {17},
pages = {7100-7104},
year = {1998},
doi = {10.1063/1.477393},
URL = { https://doi.org/10.1063/1.477393},
eprint = {https://doi.org/10.1063/1.477393}
}

@article{schlegel_1998_vibr_analisis,
    author = {Ayala, Philippe Y. and Schlegel, H. Bernhard},
    title = "{Identification and treatment of internal rotation in normal mode vibrational analysis}",
    journal = {The Journal of Chemical Physics},
    volume = {108},
    number = {6},
    pages = {2314-2325},
    year = {1998},
    month = {02},
    issn = {0021-9606},
    doi = {10.1063/1.475616},
    url = {https://doi.org/10.1063/1.475616},
    eprint = {https://pubs.aip.org/aip/jcp/article-pdf/108/6/2314/10792051/2314\_1\_online.pdf},
}

@misc{py3dmol,
  title={Py3Dmol: an IPython interface for embedding 3Dmol.js views in Jupyter notebooks},
 author={Koes, David},
year={2016},
 howpublished={\url{https://pypi.org/project/py3Dmol/}}
}

@misc{Git_HML,
	author = {Giorgio Domenichini},
	title = {Hessian Machine Learning},
	howpublished="\url{https://github.com/giorgiodomen/Hessian_Machine_Learning}",
}

@misc{scikit-learn,
 title={Scikit-learn: Machine Learning in Python},
 author={Pedregosa, F. and Varoquaux, G. and Gramfort, A. and Michel, V.
         and Thirion, B. and Grisel, O. and Blondel, M. and Prettenhofer, P.
         and Weiss, R. and Dubourg, V. and Vanderplas, J. and Passos, A. and
         Cournapeau, D. and Brucher, M. and Perrot, M. and Duchesnay, E.},
 journal={Journal of Machine Learning Research},
 volume={12},
 pages={2825--2830},
 year={2011},
 howpublished={\url{https://jmlr.org/papers/v12/pedregosa11a.html}}
}

@misc{jax2026github,
	author = {James Bradbury and Roy Frostig and Peter Hawkins and Matthew James Johnson and Chris Leary and Dougal Maclaurin and George Necula and Adam Paszke and Jake Vander Plas and Skye Wanderman-Milne and Qiao Zhang},
	title = {JAX: Gunda-accelerated machine learning research via highly composable platform-independent compiler research},
	url = {http://github.com/google/jax},
	version = {0.9.2},
	year = {2026},
}

@misc{flax2026github,
	author = {Jonathan Heek and Anselm Levskaya and Avital Oliver and Marvin Ritter and Bertrand Rondepierre and Andreas Steiner and Marc van Zee},
	title = {Flax: A neural network library and ecosystem for {JAX} designed for flexibility},
	url = {http://github.com/google/flax},
	version = {0.12.6},
	year = {2026},
}

@misc{optax2026github,
	author = {Matteo Hessel and David Budden and Fabio Viola and Mihaela Rosca and Jeroen Erenst and Tom Hennigan},
	title = {Optax: composable gradient transformation and optimization, in JAX},
	url = {http://github.com/google-deepmind/optax},
	version = {0.2.8},
	year = {2026},
}

@misc{orbax2026github,
	title = {Orbax: JAX training utilities for checkpointing and more},
	url = {http://github.com/google/orbax},
	version = {0.11.35},
	year = {2026}
}

@article{hunter2007matplotlib,
  title={Matplotlib: A 2D graphics environment},
  author={Hunter, John D},
  journal={Computing in science \& engineering},
  volume={9},
  number={03},
  pages={90--95},
  year={2007},
  publisher={IEEE Computer Society}
}

@article{   harris2020numpy,
 title         = {Array programming with {NumPy}},
 author        = {Charles R. Harris and K. Jarrod Millman and Stefan J.
                 van der Walt and Ralf Gommers and Pauli Virtanen and David
                 Cournapeau and Eric Wieser and Julian Taylor and Sebastian
                 Berg and Nathaniel J. Smith and Robert Kern and Matti Picus
                 and Stephan Hoyer and Marten H. van Kerkwijk and Matthew
                 Brett and Allan Haldane and Jaime Fernandez del
                 Rio and Mark Wiebe and Pearu Peterson and Pierre
                 Gerard-Marchant and Kevin Sheppard and Tyler Reddy and
                 Warren Weckesser and Hameer Abbasi and Christoph Gohlke and
                 Travis E. Oliphant},
 year          = {2020},
 month         = sep,
 journal       = {Nature},
 volume        = {585},
 number        = {7825},
 pages         = {357--362},
 doi           = {10.1038/s41586-020-2649-2},
 publisher     = {Springer Science and Business Media {LLC}},
 url           = {https://doi.org/10.1038/s41586-020-2649-2}
}

@ARTICLE{2020SciPy-NMeth,
  author  = {Virtanen, Pauli and Gommers, Ralf and Oliphant, Travis E. and
            Haberland, Matt and Reddy, Tyler and Cournapeau, David and
            Burovski, Evgeni and Peterson, Pearu and Weckesser, Warren and
            Bright, Jonathan and {van der Walt}, Stefan J. and
            Brett, Matthew and Wilson, Joshua and Millman, K. Jarrod and
            Mayorov, Nikolay and Nelson, Andrew R. J. and Jones, Eric and
            Kern, Robert and Larson, Eric and Carey, C J and
            Polat, Ilhan and Feng, Yu and Moore, Eric W. and
            {VanderPlas}, Jake and Laxalde, Denis and Perktold, Josef and
            Cimrman, Robert and Henriksen, Ian and Quintero, E. A. and
            Harris, Charles R. and Archibald, Anne M. and
            Ribeiro, Antonio H. and Pedregosa, Fabian and
            {van Mulbregt}, Paul and {SciPy 1.0 Contributors}},
  title   = {{{SciPy} 1.0: Fundamental Algorithms for Scientific
            Computing in Python}},
  journal = {Nature Methods},
  year    = {2020},
  volume  = {17},
  pages   = {261--272},
  adsurl  = {https://rdcu.be/b08Wh},
  doi     = {10.1038/s41592-019-0686-2},
}

@book{van1995python,
author       = {Guido van Rossum and Python Software Foundation},
title        = {The Python Tutorial},
organization = {Python Software Foundation},
year         = {2024},
url          = {https://docs.python.org/3/tutorial/},
}

@misc{python_multiprocessing,
	title        = {multiprocessing — Process-based parallelism},
	author       = {{Python Software Foundation}},
	organization = {Python Software Foundation},
	year         = {2024},
	note         = {Python 3.13 documentation},
	url={https://docs.python.org/3/library/multiprocessing.html},
	urldate      = {2026-05-07}
}

@article{ipython,
  Author    = {P\'erez, Fernando and Granger, Brian E.},
  Title     = {{IP}ython: a System for Interactive Scientific Computing},
  Journal   = {Computing in Science and Engineering},
  Volume    = {9},
  Number    = {3},
  Pages     = {21--29},
  month     = may,
  year      = 2007,
  url       = "https://ipython.org",
  ISSN      = "1521-9615",
  doi       = {10.1109/MCSE.2007.53},
  publisher = {IEEE Computer Society},
}

@misc{Geometric_ICs,
	title = {GeomeTRIC},
	howpublished = "\url{https://geometric.readthedocs.io/en/latest/how-it-works.html\#internal-coordinate-setup}",
	year = {Accessed: 2025-11-30}
}

@article{Cordero_radii,
	author ={Cordero, Beatriz and Gómez, Verónica and Platero-Prats, Ana E. and Revés, Marc and Echeverría, Jorge and Cremades, Eduard and Barragán, Flavia and Alvarez, Santiago},
	title  ={Covalent radii revisited},
	journal  ={Dalton Trans},
	year  ={2008},
	issue  ={21},
	pages  ={2832-2838},
	publisher  ={The Royal Society of Chemistry},
	doi  ={10.1039/B801115J},
	url  ={http://dx.doi.org/10.1039/B801115J}
	}

@article{pubchem_main,
	author = {Kim, Sunghwan and Chen, Jie and Cheng, Tiejun and Gindulyte, Asta and He, Jia and He, Siqian and Li, Qingliang and Shoemaker, Benjamin A and Thiessen, Paul A and Yu, Bo and Zaslavsky, Leonid and Zhang, Jian and Bolton, Evan E},
	title = {PubChem 2025 update},
	journal = {Nucleic Acids Research},
	volume = {53},
	number = {D1},
	pages = {D1516-D1525},
	year = {2025},
	month = {01},
	issn = {1362-4962},
	doi = {10.1093/nar/gkae1059},
	url = {https://doi.org/10.1093/nar/gkae1059},
}

@misc{pubchem_TTX,
author={National Center for Biotechnology Information} ,
title={PubChem Compound Summary for CID 11174599, Tetrodotoxin. },
year={2026},
howpublished="\url{https://pubchem.ncbi.nlm.nih.gov/compound/Tetrodotoxin.}"
}

@article{rcbs_paper,
	author = {Berman, Helen M. and Westbrook, John and Feng, Zukang and Gilliland, Gary and Bhat, T. N. and Weissig, Helge and Shindyalov, Ilya N. and Bourne, Philip E.},
	title = {The Protein Data Bank},
	journal = {Nucleic Acids Research},
	volume = {28},
	number = {1},
	pages = {235-242},
	year = {2000},
	month = {01},
	issn = {0305-1048},
	doi = {10.1093/nar/28.1.235},
	url = {https://doi.org/10.1093/nar/28.1.235},
	eprint = {https://academic.oup.com/nar/article-pdf/28/1/235/9895144/280235.pdf},
}

@misc{rcbs_url,	howpublished = "\url{http://www.rcsb.org/}"}

@article{rcbs_molstar,
	author = {Sehnal, David and Bittrich, Sebastian and Deshpande, Mandar and Svobodová, Radka and Berka, Karel and Bazgier, Václav and Velankar, Sameer and Burley, Stephen K and Koča, Jaroslav and Rose, Alexander S},
	title = {Mol* Viewer: modern web app for 3D visualization and analysis of large biomolecular structures},
	journal = {Nucleic Acids Research},
	volume = {49},
	number = {W1},
	pages = {W431-W437},
	year = {2021},
	month = {07},
	issn = {0305-1048},
	doi = {10.1093/nar/gkab314},
	url = {https://doi.org/10.1093/nar/gkab314}
}

@misc{wwpdb_url,	howpublished = "\url{http://www.wwpdb.org/}"}

@article{wwpdb_paper,
	author   = {Berman, Helen and Henrick, Kim and Nakamura, Haruki},
	title    = {Announcing the worldwide {Protein Data Bank}},
	journal  = {Nature Structural Biology},
	year     = {2003},
	volume   = {10},
	number   = {12},
	pages    = {980--980},
	doi      = {10.1038/nsb1203-980}
}

@misc{1GCN_rcbs,howpublished="\url{https://www.rcsb.org/structure/1GCN}"}

@article{1GCN_pdb,
	title={X-ray analysis of glucagon and its relationship to receptor binding},
	author={Sasaki, Kyoyu and Dockerill, Susan and Adamiak, Dorota A and Tickle, Ian J and Blundell, Tom},
	doi={https://doi.org/10.2210/pdb1GCN/pdb},
	year={1977},
}
\newpage
\section{Supplementary Materials of the paper "Machine learning predictions of the Hessian matrix for peptides chains and small proteins"}

\begin{table}[H]
    \centering
	\begin{tabular}{l|r}
		Aminoacid & abbreviation \\
		\hline
		Alanine & ALA \\
		Arginine & ARG \\
		Aspartic acid &ASH\\
		Asparagine &ASN\\
		Cysteine &CYS\\
		Glutamic acid &GLH\\
		Glutamine &GLN\\
		Glycine &GLY\\
		Histidine & HIS\\
		Iso-leucine &ILE\\
		Leucine &LEU\\
		Lysine &LYN\\
		Methionine &MET\\
		Phenyl-alanine&PHE\\
		Proline &PRO\\
		Serine &SER\\
		Threonine &THR\\
		Tryptophan&TRP\\
		Tyrosine&TYR\\
		Valine&VAL	\\
	\end{tabular}
    \label{AA_table}
	\caption{The 20 proteinogenic amino acids included in the dataset, with their standard abbreviations. }
\end{table}
\begin{table}[H]
\begin{tabular}{l|c|c|c|c|c|c}
	\toprule
Nr.Res. & Build ICs & RFR pred. & to CC& RFR tot. & NN pred.& NN tot.\\ \hline
10 & 0.3"  & 2.4"  & 0.0"  & 2.7"  & 0.4"  & 0.6"  \\
20 & 0.5"  & 3.1"  & 0.0"  & 3.7"  & 0.3"  & 0.9"  \\
30 & 1.0"  & 3.5"  & 0.0"  & 4.5"  & 0.4"  & 1.4"  \\
40 & 1.5"  & 4.0"  & 0.1"  & 5.6"  & 0.4"  & 2.0"  \\
50 & 2.2"  & 4.7"  & 0.1"  & 6.9"  & 0.4"  & 2.6"  \\
60 & 3.0"  & 5.5"  & 0.2"  & 8.7"  & 0.5"  & 3.6"  \\
70 & 3.9"  & 6.6"  & 0.2"  & 10.7"  & 0.5"  & 4.5"  \\
80 & 4.8"  & 7.3"  & 0.3"  & 12.4"  & 0.6"  & 5.6"  \\
90 & 5.8"  & 7.8"  & 0.4"  & 14.0"  & 0.6"  & 6.9"  \\
100 & 6.9"  & 9.1"  & 0.5"  & 16.5"  & 0.7"  & 8.4"  \\

\end{tabular}
\end{table}

\subsection{Thermal analysis}

Within the Born-Oppenheimer approximation,  \cite{Born_Oppenheimer,atkins2023atkins} it is possible to separate the motion of the nuclei and the motion of the electron, and with the further assumption that the molecular rotational constant is independent of the vibrational state we can decompose the energy, and the thermodynamical properties of the nuclear motions in their vibrational, rotational and translational parts ($H_{TOT}=H_{VIB}+H_{ROT}+H_{TR}$, $S_{TOT}=S_{VIB}+S_{ROT}+S_{TR}$,  $G_{TOT}=G_{VIB}+G_{ROT}+G_{TR}$).

The translational and rotational contribution to the enthalpy for a non-linear isolated molecule are identical to the ones that can be calculated from the kinetic theory of gases, with $R \approx 1.987 \ {\rm cal/mol/K}$ they are:

\begin{equation} \label{HrotHtr}
	\begin{aligned}
		&H_{TR}=\frac{5}{2}RT \approx 1.5\ {\rm kcal/mol} \\
		&H_{ROT}=\frac{3}{2}RT \approx 0.9 \ {\rm kcal/mol} 
	\end{aligned}
\end{equation}

Translational and rotational contributions to the entropy can be evaluated from the relation 
\begin{equation}\label{S_rel}
	S(T)=(U(T)-U(0))/T +R ln(q)
\end{equation}

So, the entropy associated with translational motions is 
$ S_{TR}=R(\frac{5}{2}+ln(q_{TR}) $, where the partition function is $ (q_{TR}=(\frac{2\pi M k_B T}{h^2})^{3/2} k_B T/P  $

And the rotational entropy, for a non linear polyatomic molecule is $S_{ROT}=R(\frac{3}{2}+ln(q_{ROT})$, the partition function is $q_{ROT}=\left((k_B T/h)^{3/2} (\frac{\pi}{ ABC})^{0.5} \right) $, where $A,B,C$ are the moments of inertia along the three principal axes of the molecule, (the eigenvalues of the inertia tensor).
Both formulas for $S_{TR}$, and $S_{ROT}$ are derived in the approximation that $ q_{TR},q_{ROT} >> 1$, so are not valid at extreme low temperatures (eg. below ~ 2°K).

In the harmonic approximation, vibrations are treated as a set of $3N-6$ harmonic oscillators with energies levels $ E^i_n=(1/2+n)\hbar \omega_i $, where $\omega_i^2$ are the positive eigenvalues of the mass-weighted Hessian matrix $H^\mathrm{MW}_{x_Ix_J} := H_{x_Ix_J}/\sqrt{m_I {m_J}}$, in this approximation the ZPVE is the sum of ground state energies $ZPVE= (1/2) \hbar \sum_i \omega_i $.
In atomic units $\hbar =1$, the Boltzmann constant $k_B \approx 3.167 10^{-6}$, and $\beta=1/k_B T$,
the partition function is $q_{VIB}=\prod_i\sum_n e^{-\beta\omega_i n}=\prod_i 1/(1-e^{-\beta\omega_i})$, and the vibrational enthalpy (derived from the relation $H= -\partial ln(q)/\partial{\beta}$)

$$ H_{VIB}=  \sum_i \frac{ \omega_i e^{-\beta\omega_i}} {1-e^{-\beta\omega_i}}  $$

Using Eq.\ref{S_rel} the vibrational entropy for one harmonic oscillator with angular frequency $\omega$ is:
\begin{equation} \label{SHO}
	S_{HO}=\frac{1}{T} \frac{\omega e^{-\beta \omega}}{1-e^{-\beta \omega}} - k_B\log (1-e^{-\beta \omega})
\end{equation}
For very small frequencies, and in the limit that they approach zeros $0^+\leftarrow\omega_i$, the entropy $S_{HO}\rightarrow \infty$ go to minus infinity, because of the term $log(1-e^{-\beta \omega_i})$. Low frequency modes (below 50 cm$^{-1}$) often exhibit a strong anharmonicity, thus the entropy calculations in harmonic approximation exceed the true values. 

A way to overcome this problem was proposed in 2012 \cite{entropy_grimme} trough the interpolation for every mode of $S_{HO}$ with the entropy of a single rigid rotor $S_{RR}$ (can be retrieved applying eq.\ref{S_rel} to the particle on a ring problem).

\begin{equation} \label{SRR}
	S_{RR}=\frac{1}{2}k_B\left[1+\log \left(\frac{2 \pi  k_B T \mu' }{\hbar^2}\right) \right]
\end{equation}

Where $\mu'= 2\left(\frac{1}{B_{av}}+ \frac{1}{\mu}\right)^{-1}$ 
is the harmonic mean between the average of the inertia moments $B_{av}=(A+B+C)/3$, and the moment associated with the vibration $\mu=\hbar/(\omega)$ .
The interpolation is built in a way such that the RR model is preferred at low frequency and the HO at higher frequency, with an equality point determined by the empirical parameter $\tau$.

\begin{equation} \label{SmRRHO}
	S_{mRRHO}= S_{HO} \frac{1}{1+(\tau/\tilde{\nu})^\alpha} 
	+S_{RR}\left(1-\frac{1}{1+(\tau/\tilde{\nu})^\alpha} \right) 
\end{equation}

In equation \ref{SmRRHO} $\tilde{\nu}$ is the wavenumber in cm$^{-1}$ associated with the angular frequency $\omega$, $\alpha,\tau$ are empirical parameters set to be equal to 4 and 50 cm$^{-1}$ respectively.\\ 

\end{document}